\documentclass[english,aps,pra,showpacs,twocolumn,groupedaddress]{revtex4}
\usepackage[T1]{fontenc}
\usepackage[latin9]{inputenc}
\usepackage{graphicx,array,tabularx}
\usepackage{amstext}
\usepackage{hyperref}
\usepackage{bm,amssymb,amsmath,verbatim,cleveref}

\makeatletter

\@ifundefined{textcolor}{}
{%
 \definecolor{BLACK}{gray}{0}
 \definecolor{WHITE}{gray}{1}
 \definecolor{RED}{rgb}{1,0,0}
 \definecolor{GREEN}{rgb}{0,1,0}
 \definecolor{BLUE}{rgb}{0,0,1}
 \definecolor{CYAN}{cmyk}{1,0,0,0}
 \definecolor{MAGENTA}{cmyk}{0,1,0,0}
 \definecolor{YELLOW}{cmyk}{0,0,1,0}
}

\makeatother

\usepackage{babel}
\usepackage{color}

\begin{document}

\title{Adaptive identification of coherent states}
\author{Markku P.V. Stenberg}
\email[]{markku.stenberg@iki.fi}
\author{Kevin Pack}
\author{Frank K. Wilhelm}
\affiliation{Theoretical Physics, Saarland University, 66123 Saarbr{\"u}cken, Germany}

\begin{abstract}
We present methods for efficient characterization of an optical coherent state $|\alpha\rangle$. We choose measurement settings adaptively and stochastically, based on data while it is collected. Our algorithm divides the estimation into two distinct steps: (i) before the first detection of a vacuum state, the probability of choosing a measurement setting is proportional to detecting vacuum with the setting, which makes using too similar measurement settings twice unlikely; and (ii) after the first detection of vacuum, we focus measurements in the region where vacuum is most likely to be detected. In step (i) [(ii)] the detection of vacuum (a photon) has a significantly larger effect on the shape of the posterior probability distribution of $\alpha$. Compared to nonadaptive schemes, our method makes the number of measurement shots required to achieve a certain level of accuracy smaller approximately by a factor proportional to the area describing the initial uncertainty of $\alpha$ in phase space. While this algorithm is not directly robust against readout errors, we make it such by introducing repeated measurements in step (i).
\end{abstract}
\pacs{42.50.Ct, 42.50.Ar, 42.50.Dv, 03.65.Wj}
\maketitle
\section{Introduction}
A fundamental task in quantum optics is the reconstruction of the state of the light field, the complete description of which is contained in the density matrix. Close to the boundary between quantum and classical regions the density matrix is conveniently studied in phase space representation, e.g., through the Wigner function \cite{wigner32}, since this allows the investigation of the transition between the two regimes. There is a wealth of techniques  \cite{smithey93,lutterbach97,lvovsky02,bertet02} to measure the Wigner function, most of which require many copies of the state and many measurement shots.
Producing many copies, however, is not always possible or efficient. In systems just crossing the classical-to-quantum boundary such as nanomechanical resonators \cite{cleland04,bochmann13}, for example, data can be so noisy and prone to drift that the required number of experimental runs with nominally identical parameters is not possible. 

Often, this laborious task of full quantum state tomography is also asking a too general or too unspecific question. In many cases, it is sufficient to approximate
the state by estimating a few parameters characterizing it. This is the case, {\it e.g.}, in the simple and {\em prima facie} classical task of measuring both the quadrature amplitude $|\alpha|$ and phase arg($\alpha$) of a weak ac signal, for example, in the microwave range. These signals are represented as 
coherent quantum states $|\alpha\rangle=e^{-\frac{|\alpha|^2}{2}}\sum_{n=0}^{\infty}\frac{1}{\sqrt{n!}}\alpha^n|n\rangle$, with $|n\rangle$ photon number
eigenstate with $n$ photons. Estimating coherent states is an important stepping stone for the characterization of more complicated quantum states of light because many such states, e.g., a Schr\"odinger cat state $|\alpha\rangle+|-\alpha\rangle$ \cite{deleglise08}, a so-called voodoo cat state 
$e^{-i\frac{\pi}{3}}|e^{-i\frac{\pi}{3}}\alpha\rangle+e^{i\frac{\pi}{3}}|e^{i\frac{\pi}{3}}\alpha\rangle-|-\alpha\rangle$ \cite{hofheinz09}, and a compass state  $|\alpha\rangle+i|i\alpha\rangle-|-\alpha\rangle-i|-i\alpha\rangle$ \cite{kirchmair13}, can be presented as superpositions of coherent states. Direct applications of quadrature measurements include, for example, cosmic microwave background detection \cite{day03} and the search for dark matter axions \cite{bradley03}. 

Estimation of an optical phase alone, at fixed coherent state amplitude \cite{giovannetti04,berry09}, is the basis of many metrological applications, e.g., in magnetometry \cite{waldherr12,nusran12,hayes14}, detection of gravitational waves \cite{caves81,abramovici92,goda08}, and clock synchronization \cite{burgh05}. Phase estimation is also an indispensable component of several algorithms in quantum information processing \cite{nielsen00}. Since it is difficult to define the concept of phase measurement for a single mode only, the usual approach is to  consider two-mode measurements in an interferometer (see Fig.~1 in \cite{berry00}). The ultimate limit set by quantum mechanics to the precision of phase measurements is due to complementarity between photon number and phase. This translates to a so-called Heisenberg limit, phase variance scaling as $\sim N^{-2}$ with $N$ the total number of photons that pass through the interferometer. 
Note that inverse phase variance describes the Fisher information (variance of the score) \cite{lehmann98} of the phase estimate.

Optimal measurements for phase estimation have been identified theoretically in \cite{holevo79,sanders95,sanders97,luis96,buzek99,imai09} but not realized experimentally; it is not possible to perform them with photodetections at the output of the interferometer. Input states and measurements with an error scaling close to optimum (but with a different prefactor) have been proposed optimizing adaptively the next measurement in the series of measurements \cite{berry00,berry01} (local optimization). Adaptive measurements have also been designed attempting to optimize the whole series of measurements \cite{fujiwara06,hentschel10,hayashi11,hentschel11} (global optimization). The input state to the interferometer considered in \cite{sanders95,sanders97,berry00,berry01,hentschel10,hentschel11}, however, is not separable between photon number eigenstates in the output arms of the interferometer and its creation is currently an open question. Beating the standard quantum limit has, however, been demonstrated using simpler entangled input states \cite{higgins07,nagata07,okamoto08,jones09}. Here, we are going beyond bare phase estimation in addressing simultaneous adaptive estimation of both phase and amplitude of $\alpha$. In terms of density operators, we thus estimate the state within a family $\rho_{\alpha}=|\alpha\rangle\hspace{-0.07cm}\langle\alpha|,\ \alpha\in {\mathbb C}$.

The most popular approach to quantum state estimation has been maximum likelihood estimation (MLE) \cite{james01}. Given measurement settings 
$S=\{s_1,\ldots,s_M\}$ and corresponding data $D=\{d_1,\ldots,d_M\}$, it seeks for a physical state $\rho$ that maximizes the likelihood functional $P(D|\rho,S)=\prod_{i=1}^{M}P(d_i|\rho,s_i)$, with $P(d_i|\rho,s_i)$ the probability to obtain the measurement outcome $d_i$ given the state $\rho$ and measurement setting $s_i$. However, MLE does not deliver confidence intervals for the estimates. Moreover, a basic problem with MLE is that it tends to assign vanishing values for certain eigenvalues of $\rho$ \cite{bloumekohout10}. This is unreasonable since it is not possible to completely rule out some measurement outcomes with a finite amount of data. 

More advanced approaches based on Bayesian inference do not suffer from these shortcomings. Bayesian inference techniques have been developed, e.g, for phase \cite{wiseman95,berry00,berry01,armen02,berry09,hayes14}, state  \cite{huszar12,kravtsov13}, and Hamiltonian \cite{schirmer09,sergeevich11,ferrie13,wiebe14a,wiebe14b,stenberg14} estimation. For certain one-parameter estimation problems it is possible to perform local Bayesian optimization of the measurement settings analytically \cite{berry00,berry01,berry09,ferrie13,wiebe14a}. For a larger number of unknown parameters, however, finding optimal measurement settings adaptively becomes generally analytically intractable. To perform the Bayesian updates numerically, a sequential Monte-Carlo approach \cite{west93,gordon93,liu00,huszar12,granade12,wiebe14a,wiebe14b,stenberg14} has recently undergone a strong development. 

Bayesian reasoning provides a general framework to assign a probability distribution for system parameters given certain data. The nature of the data, however, depends on the measurement settings chosen by the experimenter. In general it is more effective to choose the measurement settings adaptively so that they depend on the data that has been collected. Generally, the specific set of rules, also called a policy, according to which the measurement setting $s$ is to be updated, has to be developed separately for each problem at hand. 
This is true also for the recently developed technique called self-guided quantum tomography (SGQT) \cite{ferrie14}. SGQT searches the estimate of a quantum state by making measurements in the directions close to the estimate, approaching it as a power law as a function of adaptive iteration steps. The optimal prefactors in the related power laws (as well as certain coefficient added to the base of an exponential), however, depend on the problem and parameter region.
\section{Experimental setup}
For definiteness, we consider measurement of the mode using an ideal vacuum detector \cite{sperling12,govia12,govia14,oi13}. 
A vacuum detector provides a click if there is more than zero photons in the cavity but does not give any indication of the photon number beyond that. 

Such detectors have been realized in optics in the strongly coupled regime. At microwave frequencies, the recently realized \cite{chen11} Josephson photomultiplier \cite{poudel12,peropadre11} has been shown theoretically \cite{govia12} to reach this ideal vacuum detector limit at weak tunneling and long interaction time. In contrast to a standard Mach-Zender interferometric scheme for phase measurements, photon number resolving detectors or beam splitters are not needed, and there is no entanglement involved. This technology is also suited for the measurement of a qubit state and should allow better scalability to larger circuits than that based on superconducting amplifiers. The latter requires a strong auxiliary microwave pump tone that must be isolated from the qubit circuitry with bulky cryogenic  isolators, which with the former technology can be eliminated.
\begin{figure}
\includegraphics[width=0.5\textwidth]{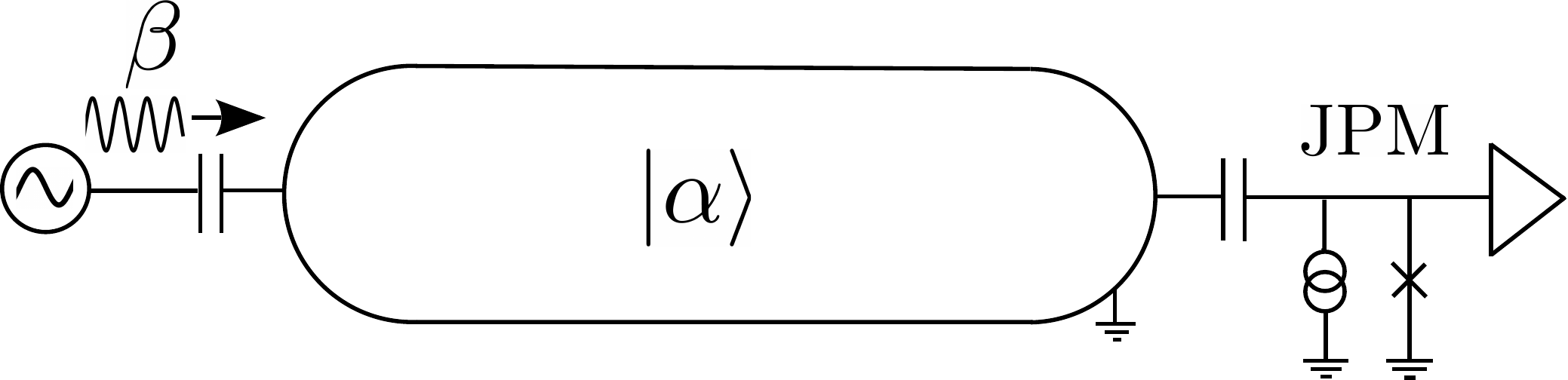}
\caption{Measurement setup. The unknown coherent state within the cavity is denoted by $|\alpha\rangle$. The microwave source emits a pulse that 
displaces the coherent state by $\beta$ in phase space. The cavity is capacitively coupled to a Josephson photomultiplier (JPM) that registers one of two possible measurement outcomes: The displaced state is either vacuum or a state containing photons.}
\label{fig:fig1}
\end{figure}

In the measurement with a vacuum detector, different points in phase space may be accessed by injecting an additional drive pulse to the input signal, cf. Fig.~\ref{fig:fig1}. 
Appropriately normalized, the drive pulse $\Omega(t)$ displaces the coherent state by an amount $\beta=-\frac{i}{2}\int\Omega(t) dt$ in phase space,
i.e., it turns $|\alpha\rangle$ into $|\alpha+\beta\rangle$, hence allowing one to measure the Husimi $Q$-function 
$Q(\alpha)=\frac{1}{\pi}\langle\alpha|\rho_{-\beta}|\alpha\rangle=\frac{1}{\pi}|\langle 0|\alpha+\beta\rangle|^2$, with $\rho_{-\beta}=|-\beta\rangle\langle-\beta|$.

In terms of a positive operator-valued measure (POVM), a measurement setting is thus characterised by a set of operators ${\mathbb M}_{\rm s}$, with $s=\beta$ given by the pulse parameter. The set ${\mathbb M}_{s}$ is defined by ${\mathbb M}_{s}=\{M_{s,d} | d\in \{\rm v,p\} \}$, where $d$ indexes the possible measurement outcomes, a vacuum state or a state containing photons. Here, we only need the operators $M_{\beta,{\rm v}}=|-\beta\rangle\langle -\beta|$ and $M_{\beta,{\rm p}}=I-M_{\beta,{\rm v}}$.

In \cite{hofheinz09} and \cite{oconnell10}, respectively, the states of a superconducting resonator and a nanomechanical resonator
were characterized. The tomographic method used in these papers is analogous to the measurement model above, since it consists of displacing the resonator state by a microwave pulse and then performing a projective measurement of the qubit. In \cite{merkel10} a nonadaptive tomographic scheme based on semidefinite programming was presented for the characterization of NOON states in resonators coupled to qubits. This type of measurements can be contrasted to probing of the Wigner function by full photon counting with number resolution \cite{banaszek96,banaszek99}.
\section{Bayesian inference}
\begin{figure} 
\includegraphics[width=0.5\textwidth]{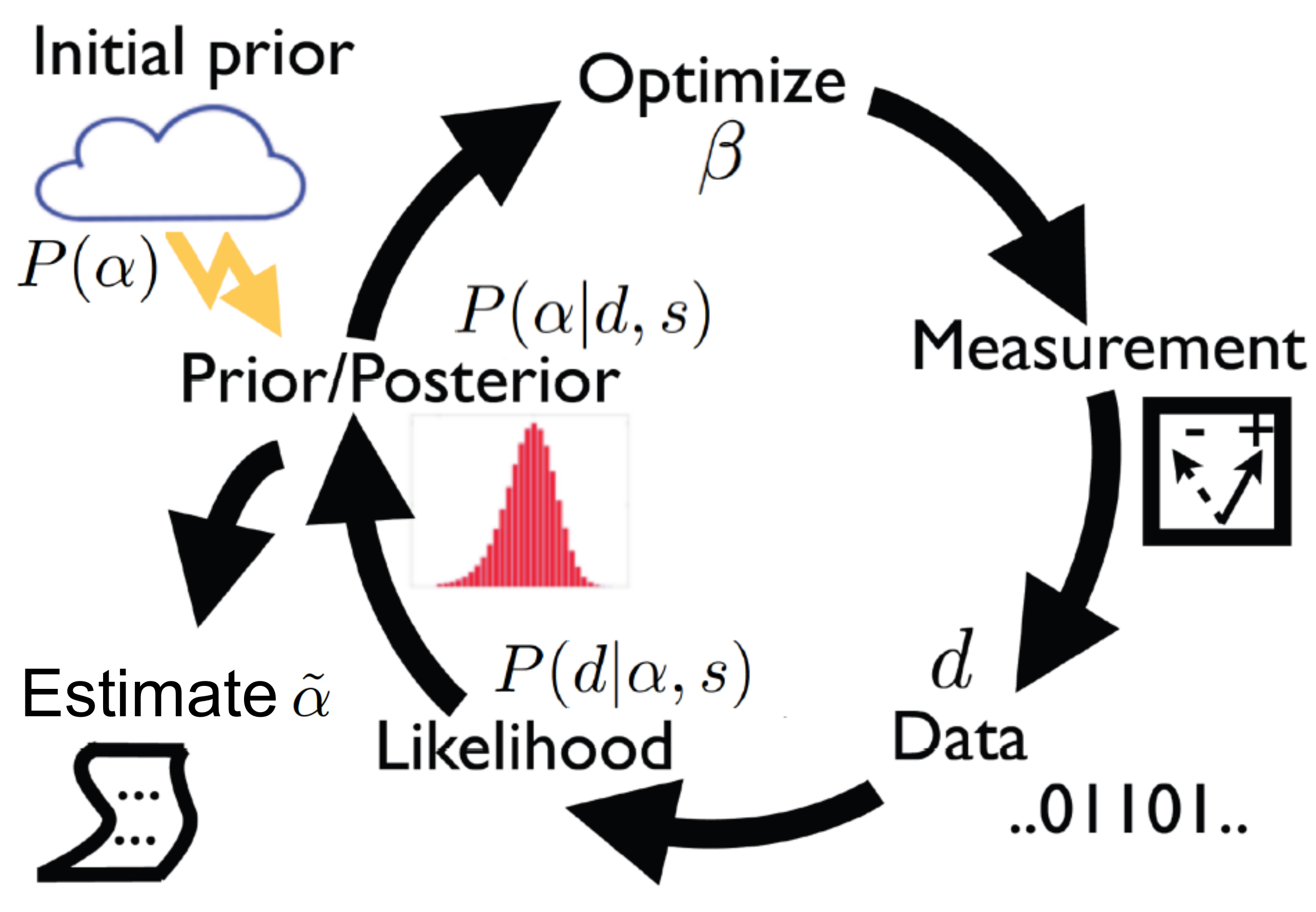}
\caption{(Color online) Illustration of an adaptive Bayesian inference scheme.}
\label{fig:fig2}
\end{figure}
The basic difference between frequentist and Bayesian approaches to parameter estimation is that the latter allow one to assign an initial prior probability distribution $P(\alpha)$ to describe an unknown state $|\alpha\rangle$. It quantifies the experimenters {\it a priori} conception of the state and its uncertainty.

Once the initial prior is set and given a suitable policy, one can take advantage of the information contained in the prior in choosing the measurement setting $s$. 
The policies used in this paper are described in Sec.~V. Bayesian inference proceeds by iteratively applying Bayes' theorem
\begin{equation}
P(\alpha|d,s)=\frac{P(d|\alpha,s)P(\alpha)}{P(d|s)}
\label{eq:bayes_theorem}
\end{equation}
as illustrated in Fig.~\ref{fig:fig2}.
Here $P(d|\alpha,s)$, referred to as likelihood, is the probability to obtain data $d$ (here, $d={\rm v,p}$) in the state $|\alpha\rangle$, given the measurement setting $s$. 
Using the notation of POVM in Sec.~II, it is related to the Hermitian operators $M_{s,d}$ through Born's rule $P(d|\alpha,s)={\rm Tr}(M_{s,d}\rho_{\alpha})$.
The normalization factor $P(d|s)$ is obtained by integrating the likelihood over all possible states $P(d|s)=\int P(d|\alpha,s)P(\alpha)d\alpha$.  The probability distribution $P(\alpha|d,s)$ for $|\alpha\rangle$ given data $d$ and measurement setting $s$ is called the posterior. The posterior can be set as the prior for the next measurement which allows iterative application of Eq.~({\ref{eq:bayes_theorem}}) 
\begin{align}
&P[\alpha|(d_{M+1},s_{M+1}),(D,S)]=\nonumber\\
&\frac{P(d_{M+1}|\alpha,s_{M+1})P[\alpha|(D,S)]}{\int P(d_{M+1}|\alpha,s_{M+1})P[\alpha|(D,S)]d\alpha},\nonumber \\
&(D,S)=\{(d_{M},s_{M})\ldots, (d_{1},s_{1})\}.
\label{eq:iterative_bayes}
\end{align}
Once a sufficient amount of data has been collected and the posterior is narrow enough, the estimate 
$\tilde{\alpha}$ is obtained from its mean value. The functional form of our likelihood function $P(d|\alpha,s)$ as well the rules to choose the measurement settings $s$ are described in Sec.~V.
\section{Numerical method}
The sequential Monte-Carlo method \cite{west93,gordon93,liu00,huszar12,granade12} delivers an efficient numerical method to perform the updates of the posterior. The posterior is approximated by keeping track of its value in  $N_{\rm p}$ moving grid points, or ``particles,'' $P[\alpha|(D,S)]\approx \sum_{n=1}^{N_{\rm p}}w_n\delta(\alpha-\alpha_n)$. Here $\alpha_n$ are the locations of the particles while $w_n$ are their relative probabilities or weights that can be updated through Bayes' theorem,
\begin{align}
&\tilde{w}_n^{(m+1)} \leftarrow P(d_{m+1}|\alpha_n,s_n)w_n^{(m)},\\
&w_n^{(m+1)} \leftarrow \frac{\tilde{w}_n^{(m+1)}}{\sum_{n=1}^{N_{\rm p}}\tilde{w}_n^{(m+1)}}.
\label{eq:w_update}
\end{align}
Here, $w_n^{(m)}$ are the weights evaluated after the $m$th measurement shot. Equation (\ref{eq:w_update}) ensures the norming $\sum_{n=1}^{N_{\rm p}}w_n=1$ and the conservation of probability.

In the following, the key quantities are the mean $\tilde{\alpha}$ and the covariance matrix ${\rm Cov}(\alpha)$
of $\alpha$ over the posterior. Numerically, these are readily approximated through 
\begin{align}
&\tilde{\alpha}=\int  P(\alpha|D)\alpha d\alpha\approx \sum_{n=1}^{N_{\rm p}}w_n\alpha_{n},\\
&{\rm Cov}({\alpha})= \int  P(\alpha|D)\alpha\alpha^T d\alpha-\tilde{\alpha}\tilde{\alpha}^T\approx \sum_{n=1}^{N_{\rm p}}w_n\alpha_{n}\alpha_{n}^T-\tilde{\alpha}\tilde{\alpha}^T,\nonumber\\
&\alpha={{\rm Re}(\alpha)\choose{{\rm Im}(\alpha)}},\quad \alpha^T=[{\rm Re}(\alpha),{\rm Im}(\alpha)].
\label{eq:mean_cov}
\end{align}

A fixed grid would limit the achievable precision of the estimate, but we make use of an adaptive grid  \cite{west93,gordon93,liu00,huszar12,granade12} 
which makes it possible to focus the particles in the regions where the probability distribution concentrates. Here, $N_{\rm p}$ locations $\alpha_n$ are first chosen following the discrete probability distribution $\{w_n\}_{n=1}^{N_{\rm p}}$. The particles are then assigned new locations $\alpha_n'$ by sampling from the normal distribution
\begin{equation}
\alpha_n'\sim \mathcal{N}[\mu_n,(1-a^2){\rm Cov}(\alpha)],
\label{eq:resampling}
\end{equation}
with the mean 
\begin{equation}
\mu_i = a\alpha_n+(1-a)\tilde{\alpha}
\label{eq:mui}
\end{equation}
and the covariance matrix $(1-a^2){\rm Cov}(\alpha)$. Here, $a$ is a parameter that we set to $a=0.999$ $95$. Finally all the weights are set to $w_n=\frac{1}{N_{\rm p}}$. The artificial dynamics induced by Eqs.~(\ref{eq:resampling}) and (\ref{eq:mui}) conserves by construction the covariance matrix ${\rm Cov}(\alpha).$

\begin{figure} 
\includegraphics[width=0.5\textwidth]{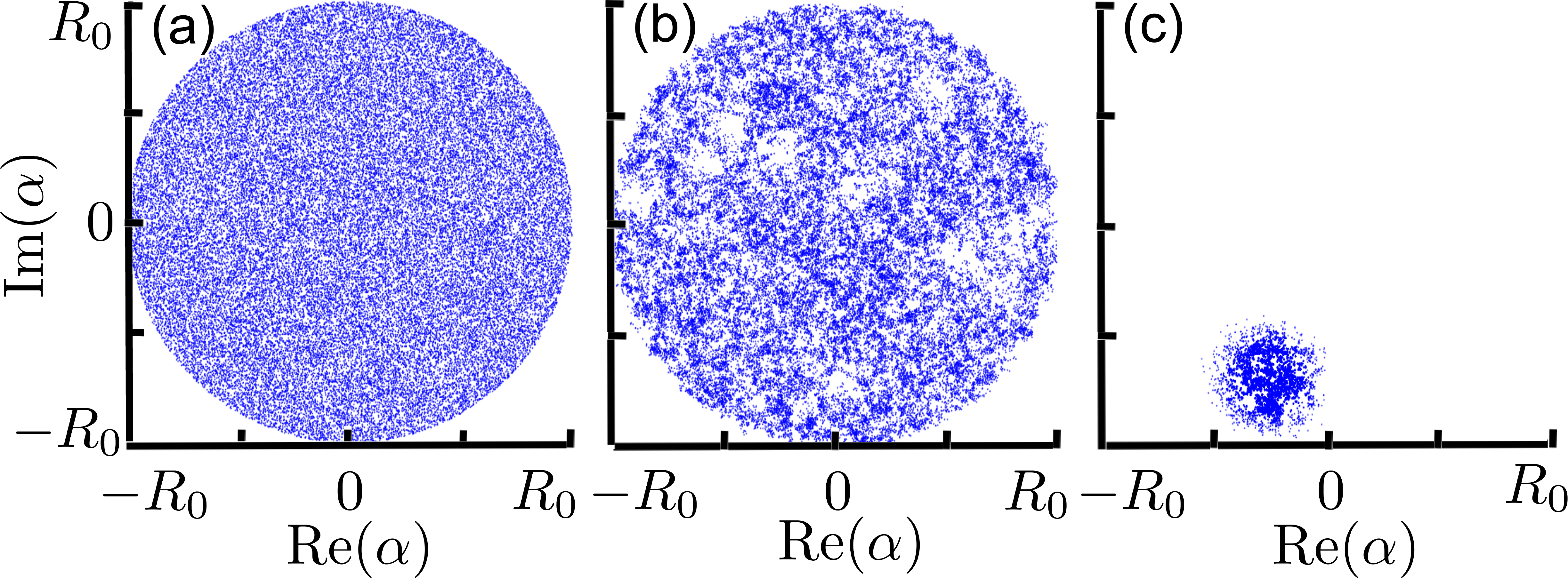}
\caption{(Color online) Exemplary evolution of the particles (see text) describing the probability distribution of the unknown coherent state $|\alpha\rangle$ in phase space. (a) Uniform initial prior within the disk $|\alpha|<R_0$. Here, $R_0=10$. (b) Posterior after ten measurement shots. The ``holes'' at positions 
$\{\alpha'\}$  have been created by photon detections with 
pulse parameters $\beta=-\alpha'$. (c) Posterior after the first detection of a vacuum state (here the 11th measurement shot). The weight of the posterior is concentrated near  $\alpha=-\beta_1$, with $\beta_1$ the pulse parameter corresponding to the first vacuum detection.}
\label{fig:fig3}
\end{figure}

In our calculations we choose $N_{\rm p}=50\ 000$. 
To compare different policies, we apply them on \mbox{15 000} simulated samples with randomly chosen true values $\alpha$.
We choose $\alpha$ from a uniform distribution on the region $|\alpha|<R_0$, with $R_0=10$ [see Fig.~\ref{fig:fig3}(a)]. Note that $R_0$ is related
to the maximum expectation value of the photon number operator through $R_0^2=\langle\hat{n}\rangle_{\rm max}$.
The initial prior was chosen to coincide with the aforementioned probability distribution. 
While here, this prior exactly incorporates what is known about estimated quantities before data collection, it has to be noted that in an actual experimental situation there is no unique and objective way to assign the initial prior, but choosing it necessarily involves certain arbitrary or subjective elements. 

\section{Policies}
Our measurement setting $s$ is defined by the pulse parameter $\beta$. We consider an ideal vacuum detector with the likelihood functions for the detections of 
vacuum (v) and a state containing photons (p), respectively
\begin{align}
&P(d={\rm v}|\alpha,\beta) = {\rm Tr}(\rho_{\alpha}M_{\beta,{\rm v}})= e^{-x^2},\ \ x = |\alpha+\beta|\nonumber \\
&P(d={\rm p}|\alpha,\beta)= {\rm Tr}(\rho_{\alpha}M_{\beta,{\rm p}})=1-P(d={\rm v}|\alpha,\beta).
\label{eq:vplikelihood}
\end{align}
In the beginning of the experiment, the first detection of a vacuum state narrows the posterior $P(\alpha|D)$ significantly more than detection of a photon. 
In the latter case the posterior only changes in the proximity of $\alpha \approx -\beta$, where its value considerably decreases [see Fig.~\ref{fig:fig3}(b)]. However, in the former case the weight of the posterior is concentrated near $\alpha \approx -\beta$, while outside this region the posterior is exponentially decreased [see Fig.~\ref{fig:fig3}(c)]. We therefore start the experiment by choosing the displacement pulse $\beta$ randomly from a probability distribution 
$P_\beta$ such that $P_{-\beta}=P[\beta|(D,S)]$ equals the posterior (here the argument $\alpha$ has to be replaced by $\beta$). 
This makes measurements with similar values of $\beta$ unlikely. After the first detection of 
vacuum we adjust the support of $P_\beta$ in the proximity of $\alpha \approx -\beta_{1}$, with $\beta_{1}$ the measurement setting with which
vacuum is detected. Before the first vacuum detection, the measurements are thus relatively uninformative, whereas most of the vacuum detections take place
within a region with a radius $\mathcal{O}(1)$ in phase space (see below). Hence, compared to nonadaptive schemes, focusing the adaptive measurements in the correct region makes the number of required measurement shots smaller approximately by a factor $\sim R_0^2$ or the area describing the initial  uncertainty of $\alpha$ in phase space. 

Specifically, we choose the measurement settings $\beta$ according to the following policy \cite{note1}
\begin{eqnarray}
P_{\beta}=
\begin{cases}
P[-\beta|(D,S)]\quad {\rm if}\ C = 0,\\
\frac{1}{\pi r^2(C)R_{\alpha}^2}\ {\rm for} \ |\beta+\tilde{\alpha}|<r(C)R_\alpha\ \ {\rm if}\ C \ge 1, \\
0\ {\rm otherwise}\quad {\rm if}\ C \ge 1.
\end{cases}
\label{eq:policy}
\end{eqnarray}
Here, $C$ is the number of measurement shots that have detected the vacuum state, $\tilde{\alpha}$ is the current mean of the posterior, and $R_\alpha$ 
describes the width of the region where the weight of the likelihood function is concentrated. More precisely, we define $R_{\alpha}=\sqrt{{\rm Tr[Cov(\alpha)]+\frac{1}{2}}}$, with ${\rm Cov}(\alpha)$ the covariance matrix for the Bayesian probability distribution of $\alpha$ [see Eq.~(\ref{eq:mean_cov})]. After the first detection of vacuum, $P_{\beta}$ is chosen to be a uniform probability distribution on a disk with a radius $r(C)R_{\alpha}$. We have carried out extensive numerical calculations to search an optimal $r(C)$ in the form of a power law
\begin{equation}
r(C)=a C^b,
\label{eq:ab}
\end{equation}
with $a$ and $b$ constants. The pair $(a,b)\in {\mathbb R}^2$ parametrizes the space within which we search for a near-optimal policy.

\begin{figure}
\includegraphics[width=0.3\textwidth]{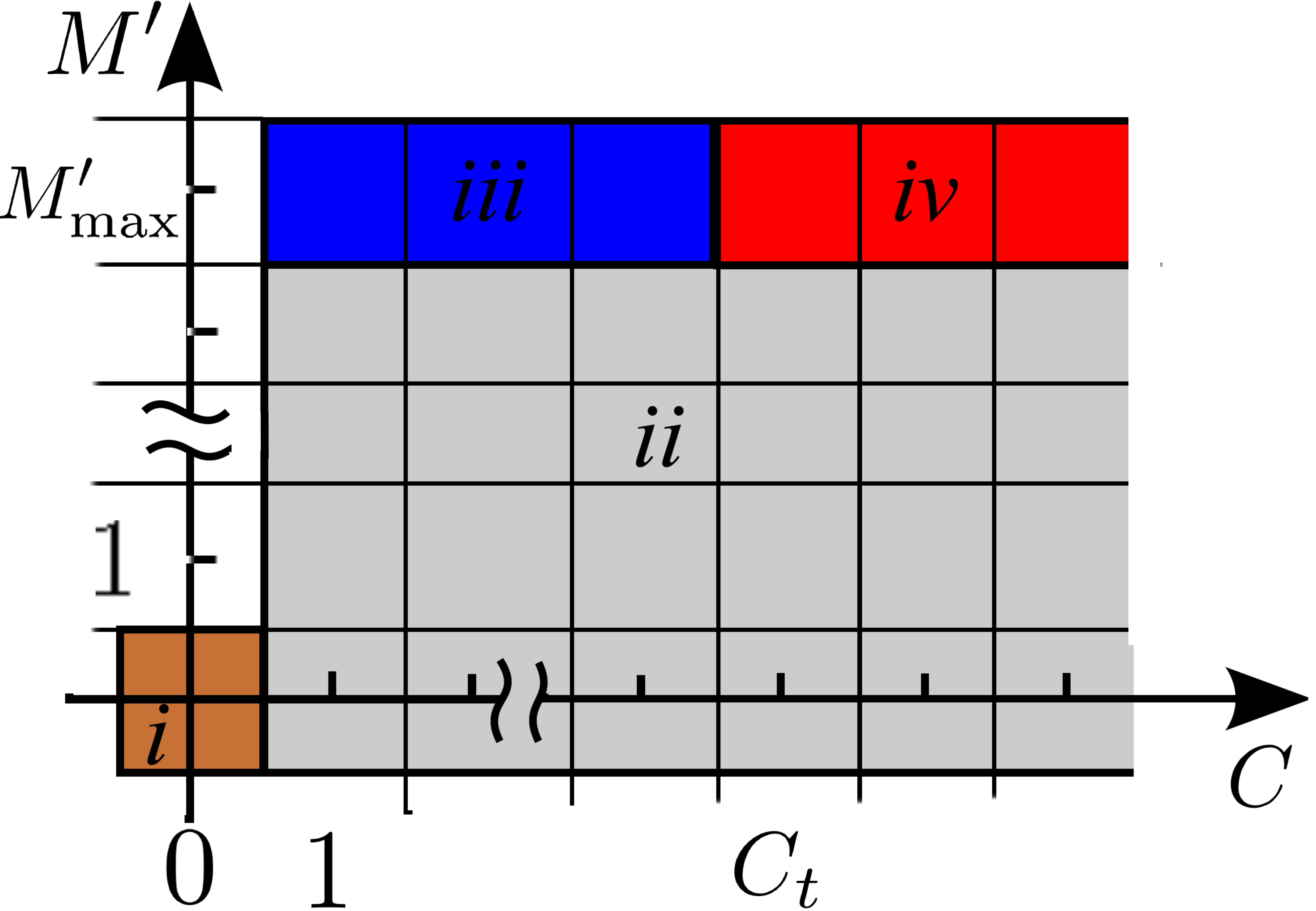}
\caption{(Color online) Illustration of the policy of Eq.~(\ref{eq:re_policy}). The vertical and horizontal axes describe the number of vacuum detections
and the number of measurement shots performed after the first vacuum detection, respectively. Different regions (i)-(iv) correspond the actions on
the first to the fourth row of Eq.~(\ref{eq:re_policy}), respectively.}
\label{fig:fig4}
\end{figure}
Since with a single vacuum detection the posterior concentrates near $\alpha \approx -\beta$, the policy (\ref{eq:policy}) is not robust against readout
errors in the experiment. However, it can be made robust against such errors by confirming that an absence of a detector click is due to vacuum state rather
than a readout error. This can be achieved through repetition. In the presence of readout errors we search policies of the form (see Fig.~\ref{fig:fig4})
\begin{eqnarray}
\begin{cases}
P_{\beta}=P[-\beta|(D,S)]\quad {\rm if}\ C = 0,\\
\beta=\beta_1,M'\rightarrow M'+1 \quad {\rm if}\ C \ge 1, M'<M'_{\rm max},\\
\beta=\beta_1, M'\rightarrow 0, C\rightarrow 0 \quad {\rm if}\ C_t > C \ge 1, M'=M'_{\rm max},\\
P_{\beta}=\frac{1}{\pi r^2(C)R_{\alpha}^2}\ {\rm for} \ |\beta+\tilde{\alpha}|<r(C)R_\alpha,\
0\ {\rm otherwise}\\ \hspace{4.5 cm} {\rm if}\ C \ge C_t, M' = M'_{\rm max}.
\end{cases}
\label{eq:re_policy}
\end{eqnarray}
Here, the measurements are repeated $M'_{\rm max}$ times at the setting $\beta_1$ that indicates vacuum state (possibly a readout error).
The variable $M'$ counts the measurement shots performed after the vacuum detection. Should after $M'_{\rm max}$ shots the number of vacuum detections 
$C$ be less than the threshold value $C_t$, the variables $M'$ and $C$ are set back to value 0. If after $M'_{\rm max}$ shots $C$ is greater or equal to the threshold value $C_t$, the remaining measurement settings are chosen as in policy (\ref{eq:policy}). Similarly as in policy (\ref{eq:policy}), we look for an optimal
policy with $r(C)$ of the form Eq.~(\ref{eq:ab}). Here, we set $M'_{\rm max}=39$, $C_{t}=15$. In our computations we assume that the probability of 
misidentifying a vacuum state as a photon state and vice versa is $P_{\rm e}=0.1$.
\section{Results}
\begin{figure}
\includegraphics[width=0.5\textwidth]{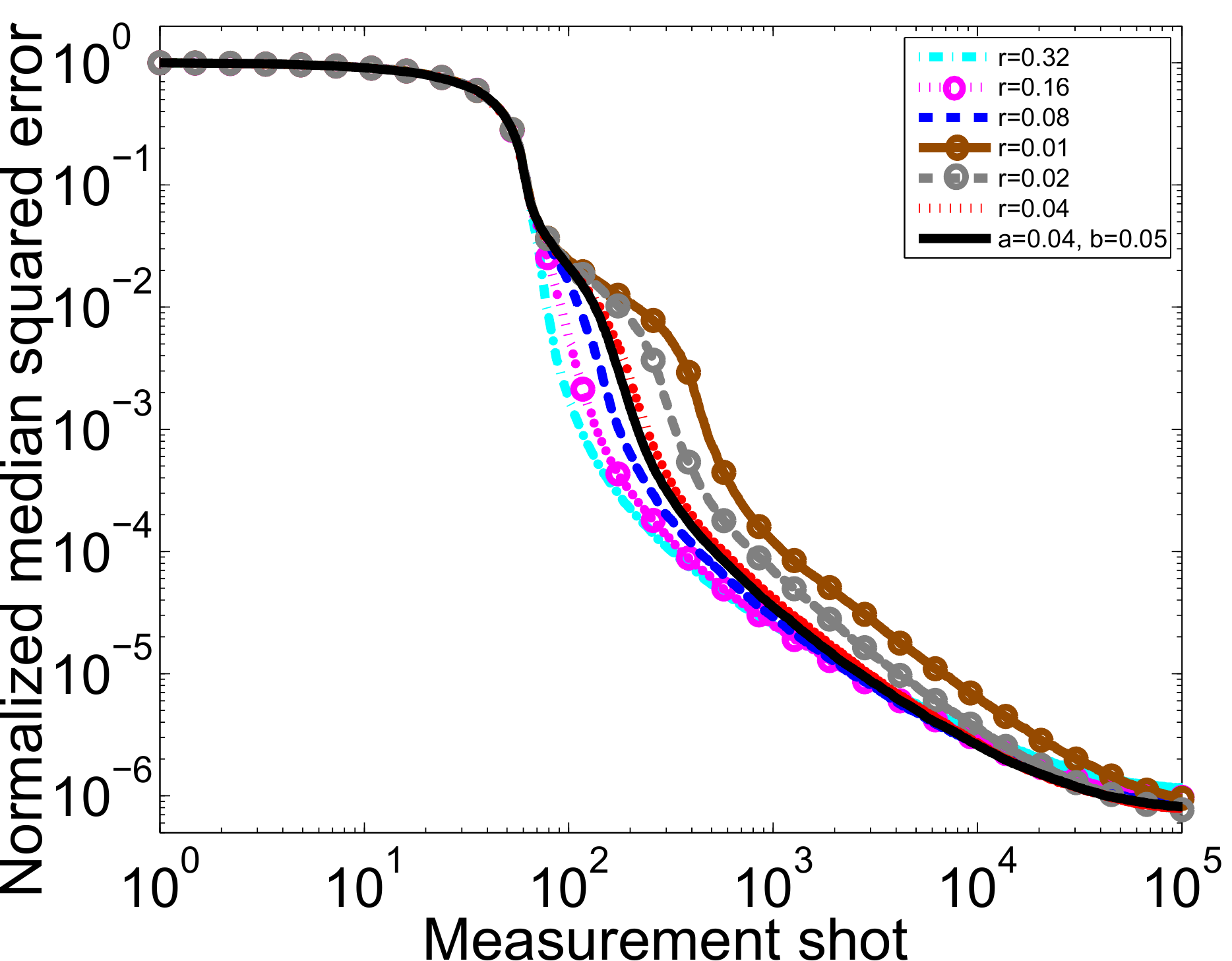}
\caption{(Color online) Median of the normalized squared error $2|\alpha_{\rm true}-\tilde{\alpha}|^2/R_0^2$ calculated from an ensemble
of $15\ 000$ simulated samples (see text) through the policies of Eq.~(\ref{eq:policy}). Different curves are for different radii $r$ denoted in the inset. Black curve is for $r(C)$ calculated through Eq.~(\ref{eq:ab}).}
\label{fig:fig5}
\end{figure}

In the absence of readout errors, we find that for an optimal policy, $a\ll 1$ and $r$ only weakly depends on $C$ (see Fig.~\ref{fig:fig5}). For instance, for the policy $\mathcal{P}$ that minimizes the median of the normalized squared error $2|\alpha_{\rm true}-\tilde{\alpha}|^2/R_0^2$ after $10^{5}$ measurement shots, we find $a=0.04$ and $b=0.05$. 
Here, the width of the plateau obtained with a low number of measurement shots depends on the degree of the initial parameter uncertainty or the radius $R_0$. By a crude estimate, 
one expects that the first detection of a vacuum state takes place after $\frac{\pi R_0^2}{\pi R_{\alpha}^2}\sim O(10^2)$ measurement shots and that the error is then rapidly
reduced to $O(1)$, corresponding to the width of the likelihood function [cf. Fig.~\ref{fig:fig3}(c)]. This is consistent with the fact that approximately after 50 measurement shots the plateau shape of the curves crosses over to a rapid decrease. Up to the level where the normalized median squared error reaches the value $\frac{1}{R_0^2}$, the different curves
overlap since until this point only the first line in Eq.~(\ref{eq:policy}) is executed, and the policy thus does not depend on $r$. The curve shape following the expected first vacuum detection (rapid decrease of the error) is universal. Interestingly, we find that the curves with different values of $r$ cross, which means that the globally best policy can not be
found by local optimization.

With $r=1$, the boundary of the disk mentioned above coincides with the steepest slope of the likelihood function $P(d|\alpha)$ of 
Eq.~(\ref{eq:vplikelihood}).
Such a disk contains approximately 39 $\%$ of the weight of $P(d={\rm v}|\alpha)$.
One might expect that policies attempting to search for the steepest slope of the likelihood function, with $a\approx 1$, would be effective, but this is not 
the case. The policies with smaller values of $a$ are able to find a crude estimate faster. In the region $x \ll 1$ the likelihood of detecting a photon $P(d={\rm p}|\alpha)\approx x^2$ is quadratically small. Therefore with $a\ll 1$, once a crude estimate has been found, most measurement shots detect vacuum and confirm the estimate. However, since the relative rate of change 
\begin{equation}
\frac{1}{P(d={\rm p}|\alpha)} \frac{\partial P(d={\rm p}|\alpha)}{\partial x} = \frac{2x}{e^{x^2}-1}\approx\frac{2}{x}
\end{equation}
increases with decreasing $x$, the rare detections of photons can effectively make a distinction between different possible values of $\alpha$
in the region $x\ll 1$. The relative rate of change above describes how much, due to Bayes' theorem (\ref{eq:bayes_theorem}), a detection of a photon
changes the relative posterior probabilities of two possible values, $\alpha$ and $\alpha'$, when $|\alpha-\alpha'|=|\Delta x|$ is fixed. Putting all together, measurement settings with $x \ll 1$ are, somewhat unexpectedly, more effective than those with larger values of $x$.

Even though for optimal policies we have $a\ll 1$, the optimal choice is not $r=0$. Indeed, the policy at $r=0$ corresponds to choosing $-\beta$ equal to the mean of the posterior, somewhat similarly with a simple, relatively ineffective, policy in the context of bare phase estimation where the control phase is chosen to coincide with the mean of the posterior (see Eq.~(6.2) in \cite{berry01}).

Policies where the measurement strategy is changed after a certain number of measurement shots have been developed for phase \cite{mitchell05} and Hamiltonian \cite{ferrie13} estimation. On the second line of Eq.~(\ref{eq:policy}), rather than on the number of all the measurement shots, we expect a possible dependence on the number of shots that take place after the first detection of a vacuum state. We therefore count in Eq.~(\ref{eq:policy}) the number of 
shots in which a vacuum state is detected.

\begin{figure}
\includegraphics[width=0.5\textwidth]{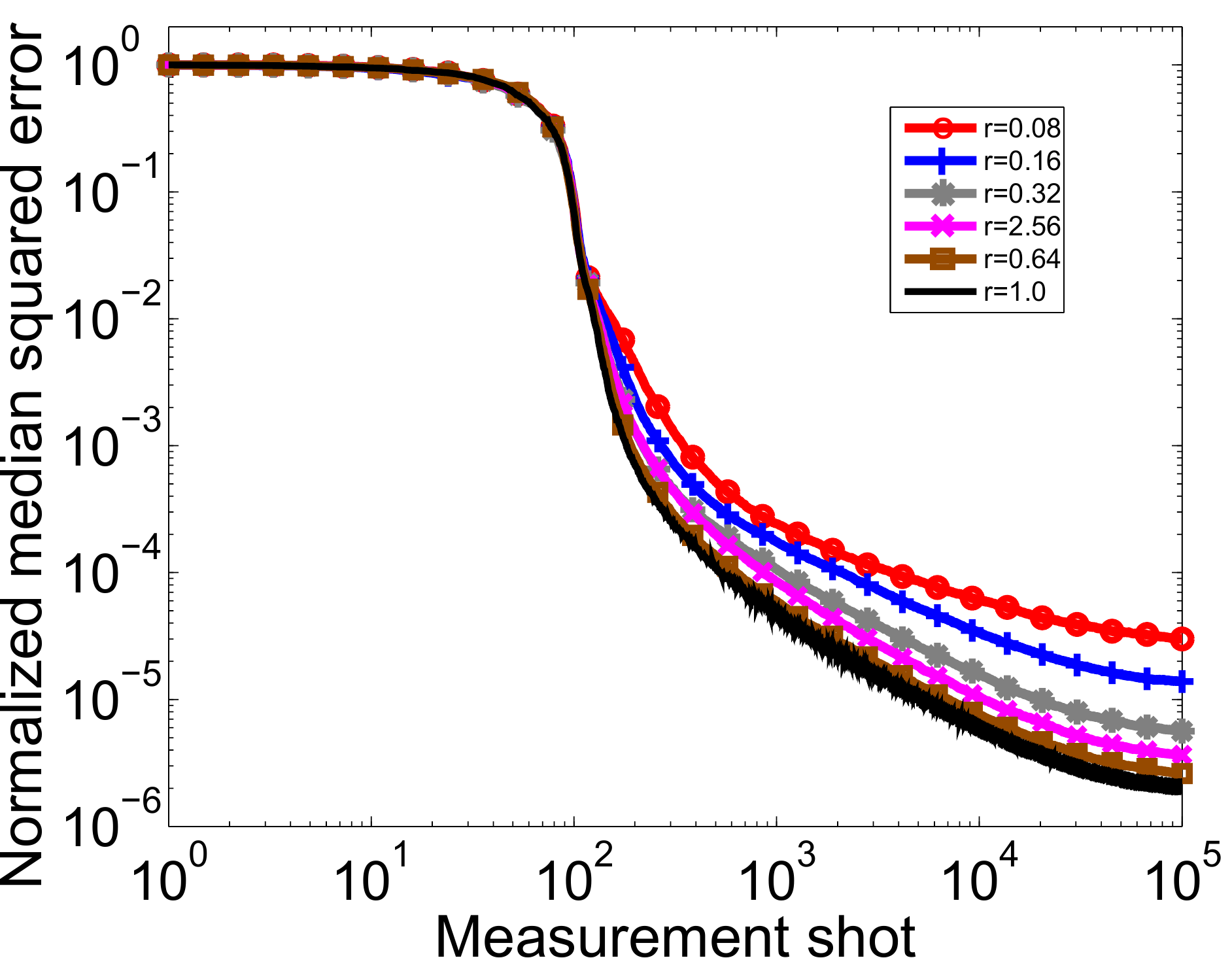}
\caption{(Color online) Median of the normalized squared error $2|\alpha_{\rm true}-\tilde{\alpha}|^2/R_0^2$ calculated from an ensemble
of $15\ 000$ simulated samples (see text) through the policies of Eq.~(\ref{eq:re_policy}) when the probability of a readout error is $P_{\rm e}=0.1$. Different curves are for different radii $r$ denoted in the
inset.}
\label{fig:fig6}
\end{figure}
In the presence of readout errors, we find that the optimal value for $a$ is larger than in the absence of these errors and the settings are therefore
more spread around their mean. The dependence of $r$ on $C$ should still be weak so that $|b| \ll 1$ (see Fig.~\ref{fig:fig6}). For the policy $\mathcal{P}_{\rm e}$ that minimizes the relative median squared error after $10^{5}$ measurement shots, we find $a=1.0$ and $b=0$.

For each simulated ensemble we obtain some samples that we refer to as ``outliers'' for which the error significantly exceeds the median and the width of the posterior probability distribution. Our policies can be made robust against such outliers through repetition as follows. After $10\ 000$ measurement shots we set the prior back to the initial prior. We thereafter perform another $10\ 000$ measurement shots. We then compare the estimates after $10\ 000$ and $20\ 000$ measurement shots. If their difference is smaller than a set threshold, we conclude that we have found a correct estimate, otherwise we start a new search of the estimate. 
For the new search we choose a prior that again coincides with the original prior. Tables I and II summarize the performance of the policies $\tilde{\mathcal{P}}$ and $\tilde{\mathcal{P}}_{\rm e}$ in the absence and presence of readout errors, respectively. These correspond to the policies $\mathcal{P}$ and $\mathcal{P}_{\rm e}$ supplemented with the outlier correction scheme. Outliers are defined as the samples with the squared error $|\tilde{\alpha}-\alpha_{\rm true}|$ larger than threshold $\mathcal{E}^2$. The outlier correction scheme appears to eliminate the outliers with an acceptable overhead.

\begin{table} 
\begin{center} 
\caption{Number of outliers per $10\ 000$ simulated samples with $P_{\rm e}=0$ for the policy $\tilde{\mathcal{P}}$ (see text). Rows correspond to the number of outliers with normalized squared error larger than $\mathcal{E}^2$ after a given number of measurement shots (indicated by the columns).}
\begin{tabular}{l  c  c  c  c  c c}
\hline\hline
$\mathcal{E}^2$/Shots & $2\times 10^4$ & $4\times 10^4$  & $6\times 10^4$ &  $8\times 10^4$ & $1.2\times 10^5$  & $1.4\times 10^5$\\ 
\hline
$10^{-5}$ & 8410 & 2062 &  504 & 118 & 5 & 0\\
$10^{-4}$ & 2218 & 173 & 3 & 0 & 0 & 0\\
$10^{-3}$ & 207 & 105 & 2 & 0 & 0 & 0\\
\hline\hline
\end{tabular} 
\label{table1} 
\end{center} 
\end{table} 

\begin{table} 
\begin{center} 
\caption{Number of outliers per $10\ 000$ simulated samples with $P_{\rm e}=0.1$ for the policy $\tilde{\mathcal{P}}_{\rm e}$ (see text). Rows correspond to the number of outliers with normalized squared error larger than $\mathcal{E}^2$ after a given number of measurement shots (indicated by the columns).}
\begin{tabular}{l  c c c c c c}
\hline\hline
$\mathcal{E}^2$/Shots & $2\times 10^4$ & $4\times 10^4$  & $6\times 10^4$ &  $8\times 10^4$ & $1.4\times 10^5$  & $2.2\times 10^5$\\ 
\hline
$10^{-5}$ & 9283 & 3474 & 1293 & 496 & 27 & 0\\
$10^{-4}$ & 5075 & 70 & 1 & 0 & 0 & 0\\
$10^{-3}$ & 825 & 42 & 0 & 0 & 0 & 0\\
\hline\hline
\end{tabular} 
\label{table1} 
\end{center} 
\end{table} 

\section{Discussion}
Based on Bayesian inference, we have delivered powerful methods for adaptive characterization of coherent states. For larger photon numbers 
$\langle \hat{n}\rangle \gg 1$, the adaptive schemes discussed here reduce the number of measurement shots required to achieve a certain level of accuracy 
by a factor proportional to the area  describing the initial uncertainty of $\alpha$ in phase space. In terms of measurement shots, we expect that efficiency can be quite generally improved by several orders of magnitude which motivates making experiments adaptively. For more complicated 
quantum states of light in a superposition $|\Psi\rangle = \sum_{n=1}^{N_{\rm s}}|\alpha_n\rangle$, our results are readily applicable for estimating $\{|\alpha_n\rangle\}_{n=1}^{N_{\rm s}}$. For full identification of $|\Psi\rangle$, our policies have to be 
supplemented with methods to obtain relative weights $\{|a_{n}|^2\}_{n=1}^{N_{\rm s}}$ of different components as well as their relative phases. Our work thus constitutes a building block that opens up an avenue for efficient estimation of multicomponent Schr\"odinger-cat states of light.
\section*{ACKNOWLEDGMENTS}
We acknowledge L. C. G. Govia, E. M. Leonard Jr., R. McDermott, I. Pechenezhskiy, G. J. Ribeill, S. F. Taylor, and T. Thorbeck for discussions. This work was supported by the European Union through ScaleQIT. 

\end{document}